\documentclass[final,3p,times]{elsarticle}

\makeatletter
\def\ps@pprintTitle{%
  \let\@oddhead\@empty
  \let\@evenhead\@empty
  \let\@oddfoot\@empty
  \let\@evenfoot\@empty
}
\makeatother

\usepackage{amssymb}
\usepackage{amsmath}
\usepackage{graphicx}
\usepackage{subcaption}
\usepackage{adjustbox}
\usepackage{lineno}
\usepackage[hidelinks]{hyperref}

\newcommand{\sqrtsnn}{\mbox{$\sqrt{s_{\mathrm{NN}}}$}}
\newcommand{\pT}{p_{\mathrm{T}}}
\newcommand{\lr}[1]{\left\langle #1\right\rangle}
\newcommand{\includefig}[2][0.30\textheight]{%
  \centering
  \adjustbox{max width=\linewidth,center}{%
    \includegraphics[height=#1,keepaspectratio]{#2}}}

\journal{Journal of Subatomic Particles and Cosmology}
\begin{document}

\begin{frontmatter}

\title{Collision energy and system size dependence of $p_{\mathrm{T}}$-differential radial flow fluctuations $v_{0}(p_{\mathrm{T}})$ at RHIC}

\author[fudan,sbu]{Zaining Wang (\textit{for the STAR Collaboration})\corref{cor1}}
\cortext[cor1]{zainingwang@foxmail.com}

\affiliation[fudan]{organization={Key Laboratory of Nuclear Physics and Ion-beam Application (MOE), and Institute of Modern Physics, Fudan University},
            city={Shanghai},
            postcode={200433},
            country={China}}
\affiliation[sbu]{organization={Department of Chemistry, Stony Brook University},
            city={Stony Brook},
            postcode={11794},
            country={USA}}

\begin{abstract}
We report the first RHIC measurements of $v_{0}(\pT)$ in Au+Au collisions at $\sqrtsnn=9.2$--$200$ GeV and O+O collisions at $\sqrtsnn=200$ GeV. The integral fluctuation $v_0$ follows a common $N_{\rm ch}$ dependence in large and small collision systems, suggesting that multiplicity is a dominant organizing variable for the overall fluctuation magnitude. The normalized response $v_0(\pT)/v_0$ exhibits an approximately common shape across the measured centralities and collision systems for $\pT\lesssim3$ GeV/$c$, consistent with an approximate factorization between the fluctuation amplitude and the $\pT$-differential spectral response. Identified-hadron mass ordering and viscous model comparisons demonstrate sensitivity to bulk viscosity ($\zeta/s$). These results establish $v_0(\pT)$ as a probe of the radial hydrodynamic response, collectivity across large and small collision systems, and QGP transport properties.
\end{abstract}


\end{frontmatter}

\section{Introduction}
\label{sec:intro}

High-energy heavy-ion collisions create a quark--gluon plasma (QGP) that undergoes collective expansion driven by strong pressure gradients
~\cite{Heinz:2013th,STAR:2017sal,Chen:2026gka}. While anisotropic flow
coefficients $v_n$ and their event-by-event fluctuations have been studied
extensively, radial collectivity and its fluctuations remain much less
constrained. Conventional probes such as $\lr{[\pT]}$, its fluctuations,
and blast-wave parameters provide primarily global information
~\cite{STAR:2024wgy,STAR:2025elk,STAR:2026vjv,Wan:2025rzg} and can be
affected by short-range non-flow, especially in small collision systems
~\cite{Saha:2025nyu}.

The recently proposed observable $v_0(\pT)$~\cite{Schenke2020,Parida2024}
correlates fractional spectral fluctuations $\delta n(\pT)$ with
$\delta[\pT]$ and provides a $\pT$-differential probe of radial-flow
fluctuations. Recent LHC measurements have revealed long-range correlations
and particle-species-dependent structures, while model calculations indicate
sensitivity to bulk viscosity~\cite{ALICE:2025iud,ATLAS:2025ztg,Jahan:2025cbp,Du2025}. 

RHIC provides unique leverage: O+O and Au+Au collisions at $\sqrtsnn=200$ GeV probe the system-size dependence at a common collision energy, while the Au+Au Beam Energy Scan tests the evolution of the response with collision energy and net-baryon density. In this work, we present STAR measurements of $v_0$ and $v_0(\pT)$ in these systems, focusing on system-size and beam-energy dependences, particle-species dependence, and sensitivity to bulk viscosity.

\section{Observables and analysis}
\label{sec:method}

The integral radial-flow fluctuation strength is defined as $v_0=\sqrt{\lr{(\delta[\pT])^2}}/\lr{[\pT]}$, where $[{\pT}]$ denotes the event-wise mean transverse momentum and $\delta[{\pT}]=[{\pT}]-\lr{[{\pT}]}$. Physically, events with smaller (larger) initial size develop stronger (weaker) radial acceleration, leading to flatter (steeper) spectra and a characteristic anti-correlation between $\delta n(\pT)$ and $\delta[{\pT}]$: negative at $\pT\lesssim\lr{[\pT]}$ and positive at $\pT\gtrsim\lr{[\pT]}$~\cite{Schenke2020,Parida2024,ATLAS:2025ztg}. Here, $\delta n(\pT)\equiv n(\pT)-\lr{n(\pT)}$.

The differential observable is
\begin{equation}
\label{eq:v0pt}
v_0(\pT)=\frac{\lr{\delta n(\pT)\,\delta[{\pT}]}}{\lr{n(\pT)}\,\lr{[\pT]}\,v_0}
\end{equation}
with $n(\pT)=N(\pT)/\int N(\pT)\,d\pT$. In the hydrodynamic picture,
$v_0$ characterizes the overall fluctuation amplitude arising from
event-by-event variations of the initial conditions. When this amplitude
approximately factorizes from the momentum-dependent response,
the normalized quantity $v_{0}(\pT)/v_0$ characterizes the normalized spectral response~\cite{Parida2024,ATLAS:2025ztg}.
A simple momentum-rescaling picture proposed in Ref.~\cite{Jia:2025radial}
provides a qualitative baseline for understanding the low-$\pT$ behavior
observed in the ATLAS measurements~\cite{ATLAS:2025ztg}.

Au+Au and O+O data at $\sqrtsnn=200$ GeV, together with Au+Au BES data at
$\sqrtsnn=9.2$, 14.5, 19.6, and 54.4 GeV, were recorded by STAR
($|\eta|<1$, $0.2<\pT<10$ GeV/$c$). Two subevents,
$-1<\eta<-\eta_{\rm gap}/2$ and $\eta_{\rm gap}/2<\eta<1$, with
$\eta_{\rm gap}=0.1$, are used to suppress short-range non-flow~\cite{Wang:2026hak}.
Systematic uncertainties are evaluated by varying event, track,
particle-identification, residual-pileup, and correlation selections,
including $\eta_{\rm gap}$ from 0.1 to 0.6. Variations of the event and track
selections give only small contributions. No significant gap dependence is observed in either system, with the uncertainty below 2\% for most $\pT$ bins and reaching about 5\% only in the highest-$\pT$ bin.
The inclusive charged-hadron $[\pT]$ entering the correlation reference is
calculated within $0.2<\pT<10$ GeV/$c$. The species-dependent
$\lr{[\pT]}_\alpha$ uses $0.2<\pT<2$ GeV/$c$ for $\pi/K$ and
$0.5<\pT<3$ GeV/$c$ for $p$.

\section{Results and discussion}
\label{sec:results}

\subsection{Mean transverse momentum fluctuations and large-to-small-system scaling}

Figure~\ref{fig1:int_and_diff_v0}(a) shows $v_0$ versus $N_{\rm ch}$ for inclusive and identified hadrons in Au+Au and O+O collisions at $\sqrtsnn=200$ GeV. The magnitude decreases toward central collisions with an approximate $1/\sqrt{N_{\rm ch}}$ trend, consistent with independent particle emission from fluctuating sub-sources. Crucially, Au+Au and O+O collisions follow a common curve at similar
$N_{\rm ch}$, which shows that the overall fluctuation
magnitude follows a similar event-activity dependence across the two collision systems. Such behavior is consistent with the expected sensitivity of radial-flow fluctuations to event-by-event variations of the initial conditions, including the transverse size. The same $N_{\rm ch}$-driven trend for mesons and baryons further supports
a common underlying fluctuation pattern. O+O collisions thus provide a direct test of whether heavy-ion-like radial flow fluctuations persist in a small system at comparable multiplicity, despite the substantially different initial geometry of the O+O system.

\subsection{Hydrodynamic response and factorization at $\sqrtsnn=200$ GeV}

\begin{figure}[!t]
\centering
\begin{subfigure}[t]{0.40\linewidth}
\includefig[0.25\textheight]{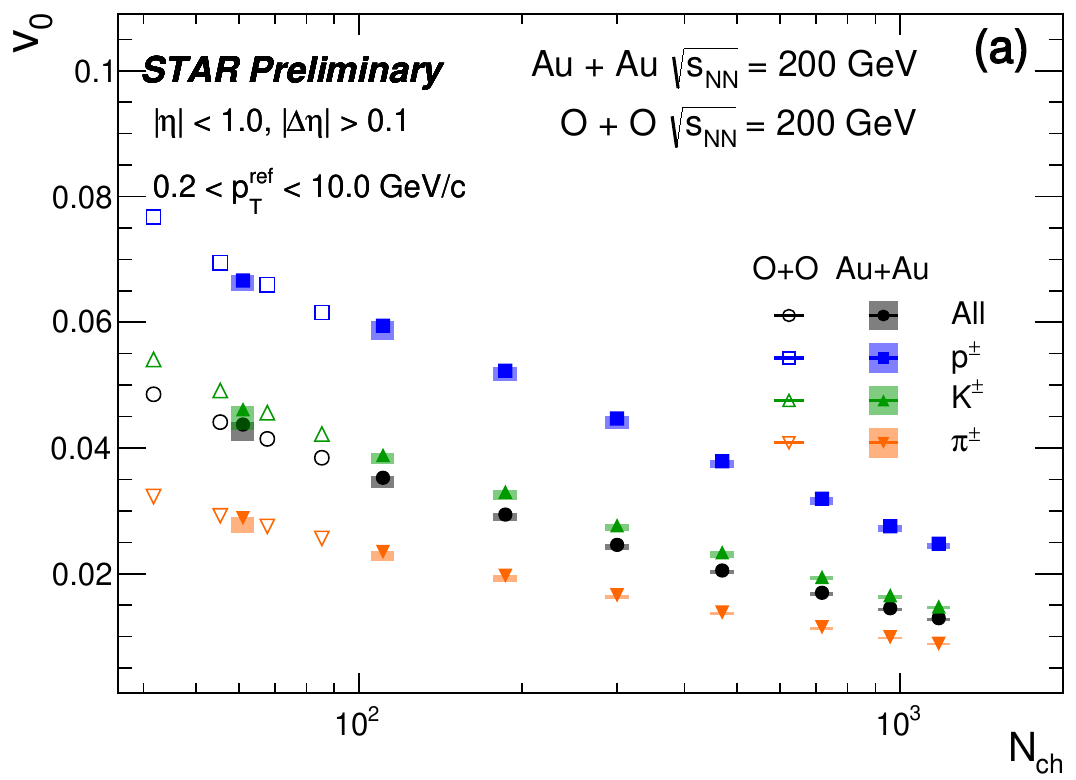}
\end{subfigure}
\hfill
\begin{subfigure}[t]{0.5\linewidth}
\includefig[0.25\textheight]{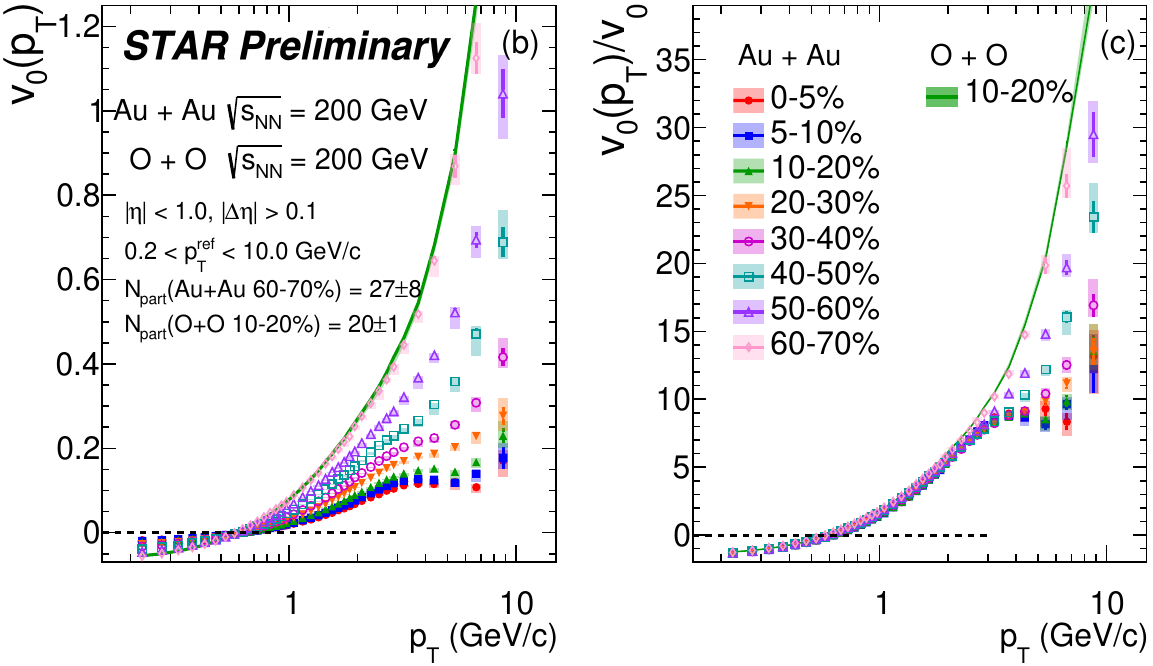}
\end{subfigure}
\caption{
Integral and differential radial-flow fluctuations in Au+Au and O+O collisions at
$\sqrtsnn=200$ GeV.
Panel (a) shows the integral fluctuation strength $v_0$ as a function of
the charged-particle multiplicity $N_{\rm ch}$ for inclusive charged hadrons
and identified $\pi^\pm$, $K^\pm$, and $p(\bar{p})$.
Panels (b) and (c) show $v_0(\pT)$ and the normalized quantity
$v_0(\pT)/v_0$ for inclusive charged hadrons in different centrality
intervals, respectively.
Reference particles are selected within $|\eta|<1$ and
$0.2<\pT^{\rm ref}<10$ GeV/$c$, with a two-subevent separation
$|\Delta\eta|>0.1$. Statistical and systematic uncertainties are represented by vertical bars and shaded boxes, respectively.
}
\label{fig1:int_and_diff_v0}
\end{figure}

Figures~\ref{fig1:int_and_diff_v0}(b) and~\ref{fig1:int_and_diff_v0}(c) show $v_0(\pT)$ and $v_0(\pT)/v_0$, respectively, for inclusive charged hadrons. While the magnitude of $v_0(\pT)$ varies strongly with centrality, the normalized ratio $v_0(\pT)/v_0$ exhibits an approximately common shape for $\pT\lesssim3$ GeV/$c$ across the measured Au+Au and O+O collisions. This behavior is consistent with a factorization picture in which the centrality dependence is carried predominantly by the overall fluctuation amplitude, while the momentum-dependent spectral response remains approximately common. The zero crossing near $\pT\approx0.6$ GeV/$c$, close to $\lr{[\pT]}$, matches spectral-slope fluctuations expected from
radial expansion~\cite{Schenke2020,Parida2024}.

\subsection{Identified hadrons and bulk-viscosity sensitivity}

\begin{figure}[!t]
\centering
\begin{subfigure}[t]{0.60\linewidth}
\includefig[0.30\textheight]{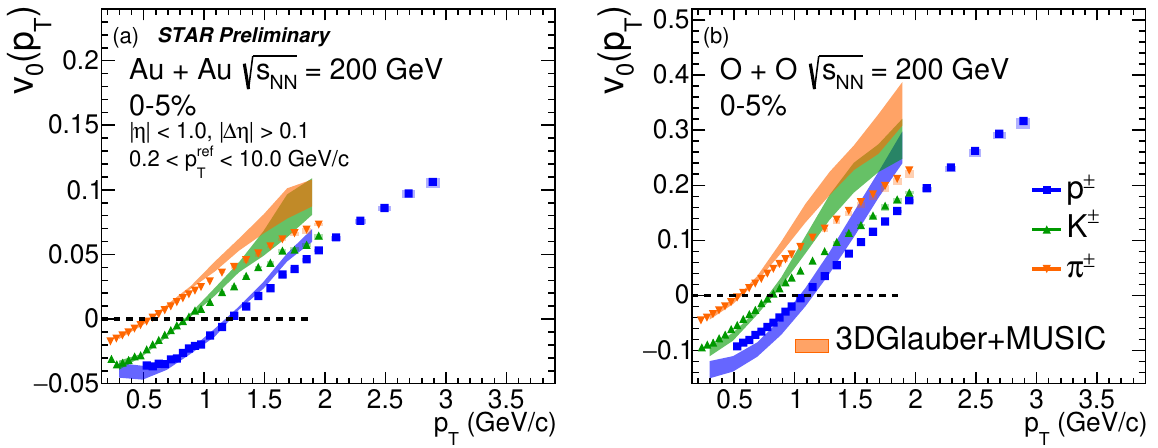}
\end{subfigure}
\hfill
\begin{subfigure}[t]{0.29\linewidth}
\includefig[0.30\textheight]{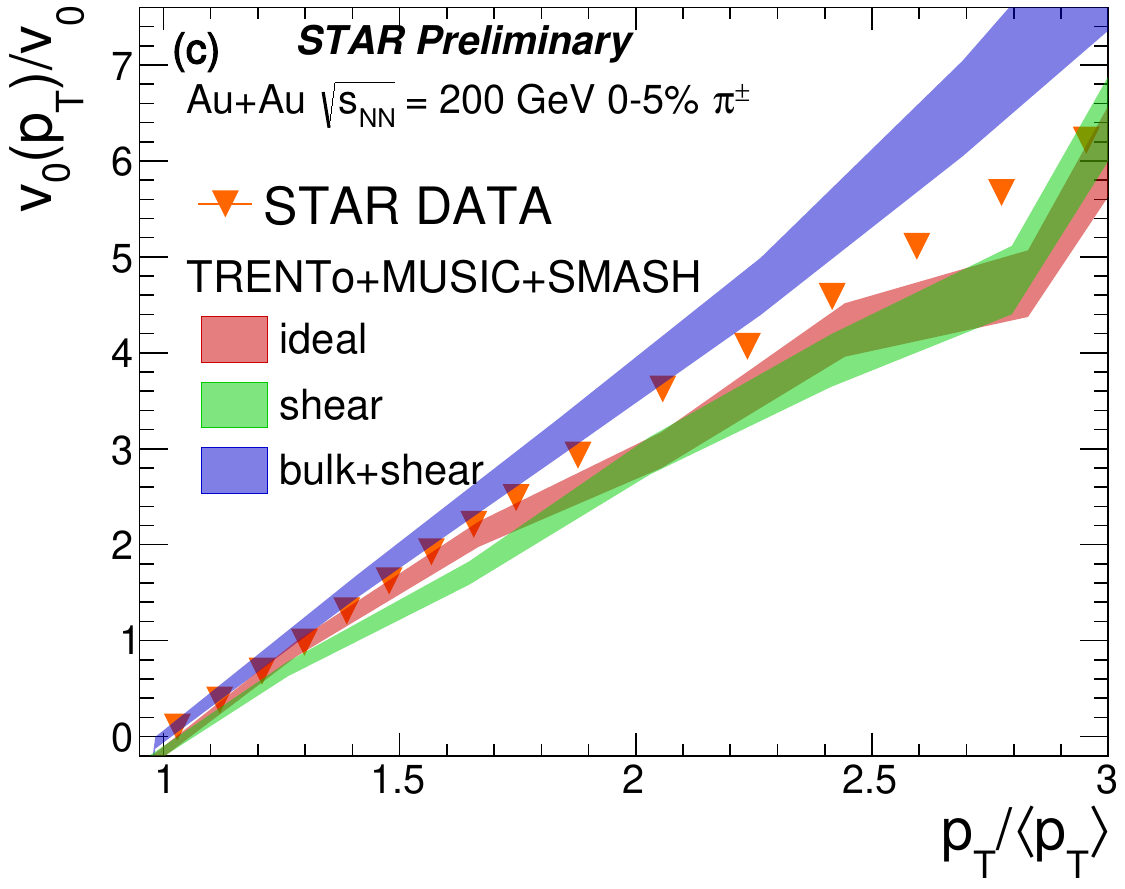}
\end{subfigure}
\caption{
Identified-hadron radial-flow fluctuations and comparisons with hydrodynamic
calculations in 0--5\% central collisions at $\sqrtsnn=200$ GeV.
Panels (a) and (b) show $v_0(\pT)$ for $\pi^\pm$, $K^\pm$, and
$p(\bar{p})$ in Au+Au and O+O collisions, respectively.
The O+O results are compared with calculations based on
3D-Glauber initial conditions followed by MUSIC hydrodynamics and a
hadronic afterburner.
Panel (c) shows the normalized charged-pion observable
$v_{0,\pi}(\pT)/v_{0,\pi}$ as a function of
$\pT/\lr{[\pT]}_{\pi}$ in Au+Au collisions,
compared with TRENTo+MUSIC+SMASH calculations with the same conventions using ideal,
shear-viscous, and shear-plus-bulk-viscous medium evolution. Statistical and systematic uncertainties are represented by vertical bars and shaded boxes, respectively.
}
\label{fig:pidmodel}
\end{figure}

Figures~\ref{fig:pidmodel}(a) and~\ref{fig:pidmodel}(b) show $v_0(\pT)$ for identified hadrons in 0--5\% Au+Au and O+O collisions. A clear mass ordering of $v_0(\pT)$ is observed at low $\pT$, with the largest magnitude for protons. For each particle species, $v_0(\pT)$ crosses zero near its corresponding $\lr{[\pT]}$. The observed mass ordering is consistent with the stronger radial boost experienced by heavier hadrons in a collectively expanding medium~\cite{Schenke2020,Parida2024}. The mass hierarchy observed in O+O collisions is similar to that in Au+Au collisions, suggesting a similar response of the particle-species-dependent transverse-momentum spectra to event-by-event radial-flow fluctuations in the two collision systems.

Bulk viscosity is expected to regulate isotropic expansion and therefore to imprint directly on radial-flow fluctuations. \texttt{3D-Glauber+MUSIC+UrQMD}~\cite{Jahan:2025cbp} reproduces the low-$\pT$ mass hierarchy in Au+Au and O+O collisions, supporting a hydrodynamic origin. For $\pi^\pm$, the \texttt{TRENTo+free-streaming+MUSIC+iSS+SMASH}~\cite{Du2025} calculation shown in Fig. 2(c) indicates that ideal and shear-only calculations underpredict the mid-$\pT$ magnitude of $v_0(\pT)/v_0$, whereas adding bulk viscosity brings the prediction closer to data. The clear separation among the viscous scenarios in the intermediate-$p_\mathrm{T}$ region makes this observable a particularly promising constraint on $\zeta/s$, one of the least constrained QGP transport coefficients.

\subsection{Beam-energy dependence}

\begin{figure}[!t]
\includefig[0.30\textheight]{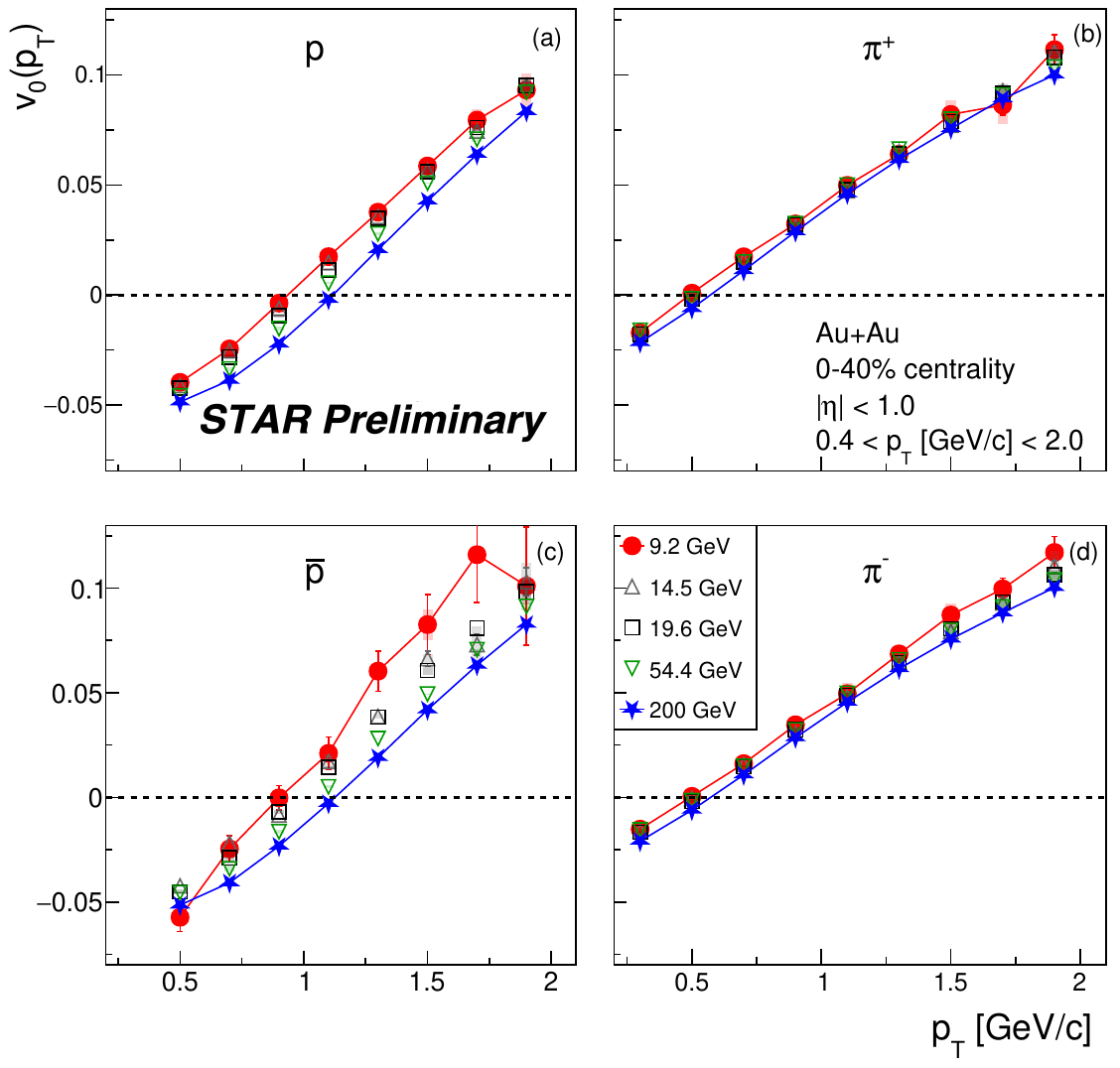}
\caption{
Beam-energy dependence of $v_0(\pT)$ in Au+Au collisions.
Panels (a)--(d) show $v_0(\pT)$ for $p$,
$\pi^+$, $\bar{p}$ and $\pi^-$ in 0--40\% Au+Au collisions at
$\sqrtsnn=9.2$, 14.5, 19.6, 54.4, and 200 GeV. Statistical and systematic uncertainties are represented by vertical bars and shaded boxes, respectively.
}
\label{fig:bes}
\end{figure}

The beam-energy dependences of $v_0(\pT)$ for pions, protons, and antiprotons are shown in Fig.~\ref{fig:bes}. The $\pi^{\pm}$ $v_0(\pT)$ exhibits only a weak beam-energy dependence, whereas the proton and antiproton $v_0(\pT)$ show a stronger variation with $\sqrtsnn$. The growing proton--antiproton difference toward lower collision energies may be qualitatively related to the increasing importance of baryon transport toward midrapidity and the associated changes in baryon spectra~\cite{STAR:2017sal}. Changes in the overall fluctuation amplitude with collision energy may also contribute to this trend.

\section{Summary}
\label{sec:summary}

We presented measurements of $v_0$ and $v_0(\pT)$ in Au+Au collisions
at $\sqrtsnn=9.2$, 14.5, 19.6, 54.4, and 200 GeV, and in O+O collisions at $\sqrtsnn=200$ GeV with the STAR detector. The common $N_{\rm ch}$ dependence of $v_0$ shows that the overall fluctuation magnitude follows a similar event-activity dependence across large and small collision systems. The approximately common low-$\pT$ shape of $v_0(\pT)/v_0$ across the measured centralities and systems is consistent with a factorization between the overall fluctuation amplitude and the momentum-dependent spectral response. Identified-hadron mass ordering and viscous model comparisons further demonstrate sensitivity to bulk viscosity, especially for pions. Together with the corresponding LHC measurements, these results further establish $v_0(\pT)$ as a robust probe of collective radial expansion and QGP transport properties.
\section*{Acknowledgments}
We thank the RHIC Operations Group and RCF at BNL for their support. This work is supported in part by the National Key Research and Development Program of China under Contract Nos. 2024YFA1612600 and 2022YFA1604900, the National Natural Science Foundation of China (NSFC) under Contract Nos. 12025501 and 12547102, and the U.S. Department of Energy, Office of Science, Office of Nuclear Physics, under Award No. DE-SC0024602.

\bibliographystyle{elsarticle-num}
\bibliography{references}

\end{document}